\documentclass[epsf,onecolumn,showpacs,preprintnumbers,prb,11pt]{revtex4-2}
\usepackage{amsmath}
\usepackage{graphicx}
\usepackage[colorlinks=true, allcolors=blue]{hyperref}
\usepackage{dcolumn}
\usepackage{bm}
\usepackage{amssymb}
\usepackage{tabularx}
\usepackage{longtable}
\usepackage[rflt]{floatflt}
\usepackage{float} 
\usepackage{tikz}
\usepackage{amsmath}
\usepackage{makeidx,amssymb,amsmath,amsthm,graphicx}
\usepackage[toc,page]{appendix}

\begin{document}
\title{Effect of symmetry reduction on the magnetic properties of LnIr$_2$Si$_2$ polymorphs}

\author{K.Kliemt$^1$}
\email{kliemt@physik.uni-frankfurt.de}
\author{J.D.Reusch$^1$}
\author{M.Bolte$^2$}
\author{C.Krellner$^1$}
\affiliation{
 $^1$Kristall- und Materiallabor, Physikalisches Institut, 
Goethe-Universit\"at Frankfurt, 60438 Frankfurt/M, Germany\\
$^2$Institut f\"ur Anorganische und Analytische Chemie, Goethe-Universit\"at Frankfurt, 60438 Frankfurt/M, Germany
}

\date{\today}

\def\neel{{N\'eel} }
\def\FA{F_{\rm an}}
\def\CA{C_{\rm an}}
\def\MS{M_{\rm sat}}
\def\EDF{E_{\rm df}  }
\def\text#1{{\rm #1}}
\def\i{\item}
\def\bv{\begin{verbatim}}
\def\ev{\end{verbatim}}
\def\ganz{Z}
\def\3{\ss}
\def\reel{{\cal}R}
\def\platz{\;\;\;\;}
\def\beginvector{\left(\begin{array}{c}   }
\def\endvector{\end{array}\right)}
\def\fff{\frac{3}{k_B} F }
\def\vec#1{ {\rm \bf #1  } }
\def\KBMUEF{\frac{3k_B}{\mu_{\rm eff}^2 }}

\begin{abstract}
	Many tetragonal compounds LnIr$_2$Si$_2$ (Ln = lanthanoid) occur in two polymorphous phases and are therefore well suited to study the relationship between crystal structure and magnetic properties. We have grown GdIr$_2$Si$_2$ single crystals of both polymorphs from a high-temperature indium flux and investigated their anisotropic magnetic properties. The higher symmetric form with the ThCr$_2$Si$_2$ structure (space group $I4/mmm$) orders antiferromagnetically at $T_{\rm N}=87\,\rm K$ while for the lower symmetric compound in the CaBe$_2$Ge$_2$ structure (space group $P4/nmm$) we determined a much lower \neel temperature $T_{\rm N}=12\,\rm K$. Our magnetic characterization of the single crystals reveals that for both compounds the magnetic moments are aligned in the $a-a$ plane of the tetragonal lattice, but that the change of the symmetry strongly effects the inplane alignment of the moment orientation. For Ln$=$Er and Ho, we confirmed the existence of LnIr$_2$Si$_2$ in the space group $P4/nmm$. The magnetic properties of these lower symmetric compounds are in remarkable difference to their related compounds in the space group $I4/mmm$. 
\end{abstract}
\maketitle





\section{Introduction}

\begin{figure}
\centering
\includegraphics[width=0.5\textwidth]{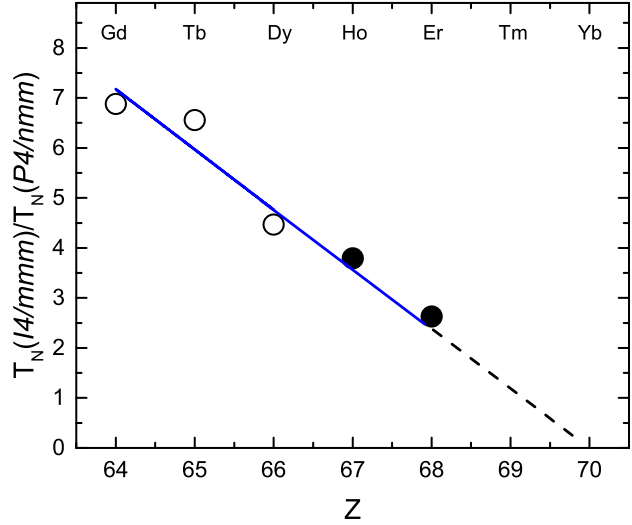}
	\caption{The ratio of the magnetic ordering temperatures of Gd(Tb,Dy)Ir$_2$Si$_2$ of the $I4/mmm$ and $P4/nmm$ phases \cite{Kliemt2019, Hirjak1984, Shigeoka2015, Sanchez1993, Uchima2017} is plotted versus the atomic number Z. The blue line is a linear fit to these data which was extended by the dashed line. This ratio calculated for HoIr$_2$Si$_2$, $P4/nmm$, with the ordering temperature $T_{\rm N}=5.8\,\rm K$ together with $T_{\rm N}=22\,\rm K$ \cite{Kliemt2018} for HoIr$_2$Si$_2$, $I4/mmm$, matches well with the dashed line (black circle).}
\label{relation}
\end{figure}

Intermetallic compounds with lanthanoids (Ln) have attracted strong research interest for several decades
due to the extraordinarily wide range of their physical properties such as superconductivity, quantum critical behavior or complex magnetic order \cite{Braun1983, Hossain2005, Shigeoka2011, Schuberth2016}. More recently Ln compounds are studied to explore their potential as data storage media as they can host skyrmions \cite{Kakihana2019} or as their spin state can be controlled and manipulated using optical fast excitations or simply by changing the temperature \cite{Windsor2020, Generalov2017}.
 In particular, the ternary compounds LnT$_2$Si$_2$ (Ln = lanthanide, T = Rh, Ir) have been investigated extensively, due to their layered structure which makes them well suited for systematic spectroscopic investigations and theoretical modeling  \cite{Guettler2016, Generalov2017, Poelchen2020, Schulz2019, Usachov2020, Vyazovskaya2019}. Compounds containing heavy elements, like Ir, are of special interest for studies of the effects of strong spin-orbit coupling on surface states \cite{Schulz2019, Usachov2020, Schulz2021}.
Among the LnIr$_2$Si$_2$ materials, polymorphism  was found for a number of compounds, like for instance YbIr$_2$Si$_2$, LaIr$_2$Si$_2$, CeIr$_2$Si$_2$, GdIr$_2$Si$_2$ and HoIr$_2$Si$_2$ \cite{Villars1991, Niepmann2001, Parthe1984}.
For these 122 compounds, two polymorphous phases are known to exist namely an I-type phase with the tetragonal body-centered ThCr$_2$Si$_2$-type structure (space group No.139, $I4/mmm$) and a P-type phase with the primitive CaBe$_2$Ge$_2$-type structure (space group No.129, $P4/nmm$).
Both structure types can be deduced from the BaAl$_4$ structure, but in the P-type phase the arrangement of the layers of Ir and Si is changed which leads to a reduced symmetry in the crystal structure.
The ThCr$_2$Si$_2$ structure type is the more common one for the 122 stoichiometry and was called the "perovskite" of intermetallics \cite{Shatruk2019}. 
Recently, a member of the CaBe$_2$Ge$_2$ structure type, CeRh$_2$As$_2$, received high attention due to very unconventional superconducting properties related to the reduced symmetry of the local environment of the {\it 4f} electrons \cite{Khim2021}.
For those systems, where both phases exist, one would expect that they show different physical properties. Therefore, single crystals of these systems are very well suited for the study of structure-property relationships.
In the past, polycrystalline GdIr$_2$Si$_2$ of both structures types has been synthesized \cite{Slaski1982, Buschow1986}, but so far the anisotropic magnetic properties of these compounds are poorly documented due to a lack of single crystals.  Single crystals became available only recently and their thermodynamic and electrical transport properties were reported in \cite{Kliemt2019}.
A first characterization of the magnetic properties of polycrystalline GdIr$_2$Si$_2$ (I-type phase) was reported by Czjzek {\it et al.} \cite{Czjzek1989} and Tung {\it et al.} \cite{Tung1997} but information on the magnetic structure of these materials are missing. It has to be mentioned that the determination of the magnetic structure of Gd compounds by neutron scattering usually is not possible since the absorption cross-section of Gd for neutrons is high. Here we present the characterization of the anisotropic magnetic properties of two polymorphs of GdIr$_2$Si$_2$ as well as implications concerning their magnetic structure.
Some of the rare-earth iridium 122 compounds have already been structurally and magnetically characterized in the I- and P-type phases. We found that the existing data on magnetic ordering temperatures for the heavy rare earths (Ln$=$Gd-Dy), open circles in Fig.~\ref{relation}, with magnetic ordering temperatures $T_{\rm N}(I4/mmm)$, $T_{\rm N}(P4/nmm)$ and the atomic number $Z$ could be used to estimate the magnetic ordering temperatures $T_{\rm N}$ for two previously non-magnetically characterized compounds HoIr$_2$Si$_2$ and ErIr$_2$Si$_2$ (closed circles) and to subsequently verify them experimentally. 
We investigated the heat capacity and the magnetic susceptibility of the up to now less studied HoIr$_2$Si$_2$ ($P4/nmm$) and compared it to that of the higher symmetric compound in the space group $I4/mmm$ \cite{Kliemt2018}. We demonstrated that the so far unknown ErIr$_2$Si$_2$ ($P4/nmm$) exists in this lower symmetric space group as well and studied its thermodynamic and magnetic properties. 

\begin{figure*}
\centering
\includegraphics[width=0.9\textwidth]{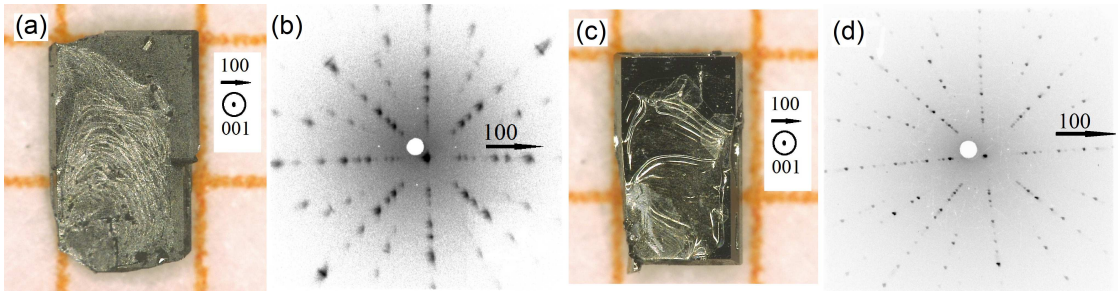}
	\caption{GdIr$_2$Si$_2$ single crystals ($I4/mmm$ (a) and $P4/nmm$ (c)) and Laue pattern ($I4/mmm$ (b) and $P4/nmm$ (d)) showing the fourfold symmetry of the tetragonal lattice.}
\label{GdIr2Si2_kristalle}
\end{figure*}

\section{Experiment}
\subsection{Crystal growth}

Single crystals of GdIr$_2$Si$_2$ were grown in indium flux following the route described earlier \cite{Kliemt2019}. 
The high purity 
starting materials Gd (99.9\%, Johnson Matthey), Ir (99.96\%, Heraeus), 
Si (99.9999\%, Wacker) 
and In (99.9995\%, Schuckard) were weighed 
in a graphite crucible and sealed in a niobium crucible 
under argon atmosphere (99.999\%). 
The elements Gd, Ir, Si and In in the ratio of Gd : Ir : Si : In = 1 : 2 : 2 : 49 were used to prepare crystals in the I-type phase. 
For CeIr$_2$Si$_2$, the P-type phase can be transformed into the I-type phase by annealing, as reported  by Niepmann and P\"ottgen \cite{Niepmann2001}.
For GdIr$_2$Si$_2$, the crystals were grown in the I-type and P-type phases directly after optimizing the initial stoichiometry. 
We found that crystals of the P-type phase were obtained for an initial stoichiometry of Gd : Ir : Si : In = 1.1 : 0.9 : 0.9 : 24. A similar route was proposed for the growth of YbIr$_2$Si$_2$ polymorphs by Hossain {\it et al.} \cite{Hossain2005}.
All of our growth experiments were done in a movable high-temperature Bridgman furnace (GERO HTRV 70-250/18). 
The maximum temperature of the furnace was $T_{\rm max}=1550^{\circ}$C which was lowered with a cooling rate of $(1 - 4)\rm K/h$ down to $T=1090^{\circ}\rm C$. Afterwards, the furnace was cooled down to room temperature with $50\,\rm K/h$. 
The indium flux was removed by etching using hydrochloric acid.
This growth procedure yielded single crystal platelets of up to (2$\times$2)\,mm$^2$, with a thickness of up to $200\,\mu\rm m$, and with masses up to $20\,\rm mg$ for the I-type phase. The crystals of the P-type phase had dimensions of (1$\times$1.5)\,mm$^2$ and a thickness of up to $300\,\mu\rm m$ with masses up to $7\,\rm mg$. Figs.~\ref{GdIr2Si2_kristalle} a) and c), show typical single crystals.
In the case of HoIr$_2$Si$_2$, our attempts to grow single crystals of the high-temperature phase using the In-flux method as described above failed.
Instead arc-melting of HoIr$_2$Si$_2$ single crystals (space group $I4/mmm$ \cite{Kliemt2018}) with subsequent fast cooling down to room temperature was used to synthesize polycrystalline samples of HoIr$_2$Si$_2$ with the space group $P4/nmm$. 
Polycrystalline samples of ErIr$_2$Si$_2$ ($P4/nmm$) were prepared by arc-melting using pieces of Er (99.9\%, ChemPur), Ir powder (99.96\%, Heraeus) and Si pieces (99.9999\%, Wacker). The samples were quickly cooled down to room temperature.

\begin{figure*}
\centering
\includegraphics[width=0.65\textwidth]{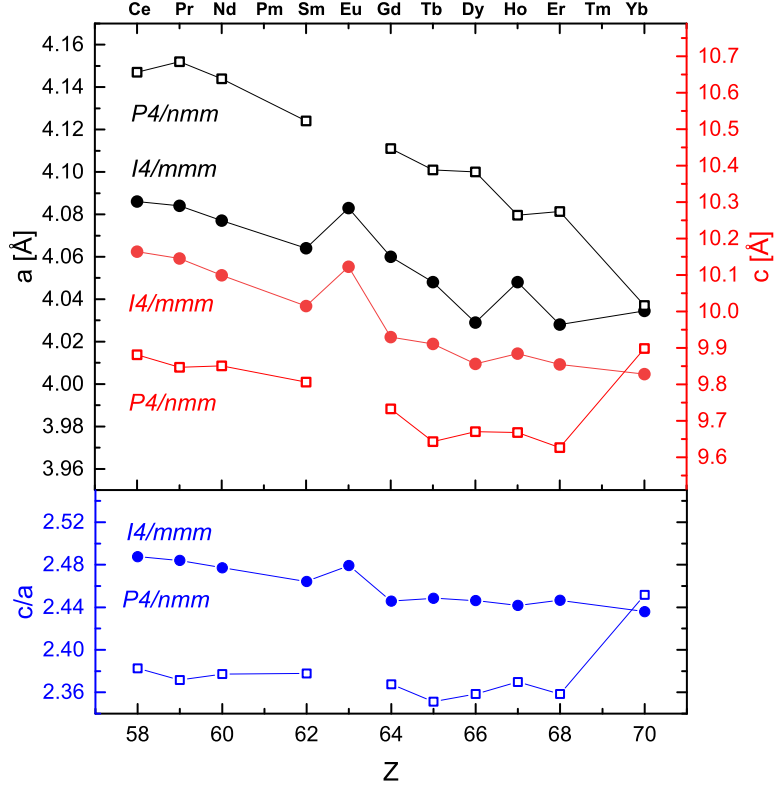}
	\caption[]{Lattice parameters of LnIr$_2$Si$_2$ $I4/mmm$ (closed circles) and $P4/nmm$ (open squares) \cite{Mihalik2009, Zhong1985, Mihalik2010, Welter2003, Valiska2013, Chevalier1986, Kliemt2019, Hirjak1984, Shigeoka2015, Melamud1984, Sanchez1993, Uchima2017, Kliemt2018, Leciejewicz1985, Krellner2012}.}
\label{LatticeParametersBild}
\end{figure*}

\subsection{Characterization}
Powder X-ray diffraction (PXRD) was performed
with a Bruker D8 diffractometer with CuK$_{\alpha}$ radiation ($\lambda = 1.5406$\,\AA) at room temperature.
PXRD on crushed single crystals of GdIr$_2$Si$_2$ in the low temperature phase 
confirmed the $I4/mmm$ tetragonal structure and yielded the lattice parameters $a=4.057$\,\AA\, and $c=9.971$\,\AA\, which are in good agreement with literature \cite{Slaski1982,Buschow1986}.
PXRD on single crystals of the high-temperature phase confirmed the $P4/nmm$ tetragonal structure and yielded the lattice parameters $a=4.100$\,\AA\, and $c=9.810$\,\AA\ in good agreement with \cite{Zhong1985}. 
While the diffraction data showed that I-type crystals are single phase, the PXRD data of the P-type phase give clear evidence that the P-type crystals are partially transformed into the I-type phase by about 30-35\%. For the magnetic characterization of the P phase, we selected large single crystals and didn't find any signatures in the magnetization or heat capacity data for the presence of I phase in these crystals.
Upon performing the structural analysis of our single crystals we found that those of the high-temperature phase (P phase) are covered by a very thin layer of the I phase which forms at a late stage of the growth process at lower temperatures. This layer can be removed by polishing the crystals and has a thickness of a few micrometers. It can be detected probing the $(001)$ surface of a P-phase crystal using a powder diffractometer in reflection mode. In the diffraction pattern of an unpolished surface only even $(00l)$ reflexes stemming from the I-phase layer are visible. After polishing the surface the odd $(00l)$ reflexes of the P phase appear. The thin I-phase layer doesn’t leave signatures in the magnetization not even of unpolished crystals. The magnetic characterization which we present in this manuscript was performed using platelet-like single crystals with an area of $(1.5\times 1.5)$mm$^2$ to $(2\times 3)$mm$^2$ while the powder data were obtained by crushing much smaller crystals of about $(0.1-0.3)$mm$^2$. The physical characterization was done on $\approx 15$ different crystals. None of the magnetization data obtained for this study nor the resistivity nor heat capacity data presented earlier in \cite{Kliemt2019} showed any signs for the presence of the I phase in the P-type crystals. Traces of the I phase were visible in the diffraction data only. We found that smaller crystallites show a higher I-phase fraction due to the higher surface to volume ratio.
Energy dispersive X-ray spectroscopy (EDX) was done with an AMETEK EDAX
Quanta400 Detector in a Zeiss scanning electron microscope (SEM) DSM 940A.
The analysis of the chemical composition by EDX microprobe analysis revealed $(20\pm 1)$ at\% Gd, $(40\pm 1)$ at\% Ir and $(40\pm 1)$ at\% Si. 
The orientation of the platelet shaped single crystals was determined 
using a Laue device, M\"uller Micro 91, with tungsten anode. 
The analysis of 10 samples yielded that the dimension perpendicular to the surface is the $[001]$ direction of the tetragonal lattice. The largest naturally grown edges point towards the $[100]$ direction of the I-type phase as well as of the P-type phase as shown in Fig.~\ref{GdIr2Si2_kristalle}
 b) and d) \cite{Kliemt2019}.
PXRD showed that the arc melted polycrystalline HoIr$_2$Si$_2$ sample solidified in the CaBe$_2$Ge$_2$-type structure. The Rietveld analysis yielded lattice parameters of $a = 4.0796$\,\AA\, and $c = 9.6677$\,\AA . It also showed that despite the fast cooling a fraction of 14\% crystallized in the ThCr$_2$Si$_2$-type structure.
The evaluation of the PXRD data showed that ErIr$_2$Si$_2$ exists in the CaBe$_2$Ge$_2$-type structure ("high-temperature phase"). Rietveld analysis revealed that a fraction of 27\% of the material crystallized in the ThCr$_2$Si$_2$-type structure ("low-temperature phase").
Single crystal structure determination on a small ErIr$_2$Si$_2$ single crystal extracted from the polycrystalline batch confirmed the space group $P4/nmm$ and yielded lattice parameters $a = 4.0666(13)$\,\AA\, and $c=9.583(4)$\,\AA.
 The data were collected on a STOE IPDS II two-circle diffractometer with a Genix Microfocus tube with mirror optics using MoK$_{\alpha}$ radiation ($\lambda = 0.71073$\,\AA). The data were scaled using the frame scaling procedure in the X-AREA program system \cite{Stoe2002}. 
 The structure was solved by direct methods using the program SHELXS \cite{Sheldrick2008} and refined against $F^2$ with full-matrix least-squares techniques using the program SHELXL \cite{Sheldrick2008}. 
 The lattice parameters and space group could be unequivocally determined. However, this method of crystallization in order to get the high temperature phase of this compound yielded only crystals of minor quality. As a result of that, the figures of merit are relatively high and significant residual electron density remains in the vicinity of the Ir and Er atoms. Fig.~\ref{LatticeParametersBild} shows the trend in the lattice parameters of the LnIr$_2$Si$_2$ compounds for both structure types. Besides the effect of the lanthanoid contraction, we find that for the P-type phase the $a$ parameter is larger compared to that of the I phase and the $c$ parameter is smaller resulting in a $c/a$ ratio of $\approx 2.35$ for the P-type phase in comparison to $c/a\approx 2.45$ for the I-type phase compounds.
After the chemical and structural characterization, we used a Quantum Design Physical Property Measurement System (PPMS) to perform magnetization measurements between $1.8\,\rm K$ and $380\,\rm K$.  
The sample was cooled in zero field prior to each measurement.

\section{Results and discussion}

\begin{figure*}
\centering
\includegraphics[width=1.0\textwidth]{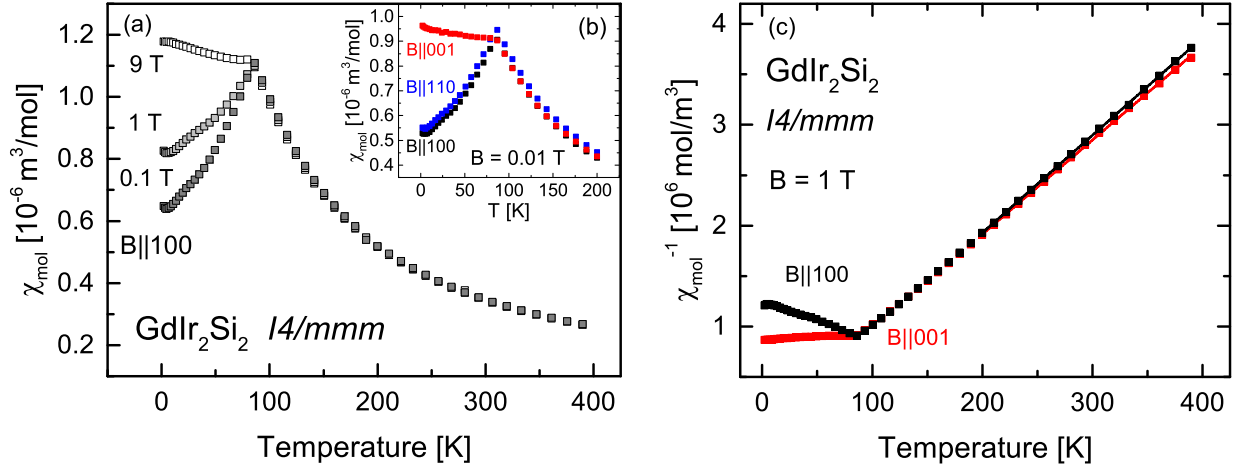}
	\caption[GdIr$_2$Si$_2$ ($I4/mmm$): Susceptibility $\chi(T)$ and inverse susceptibility $\chi(T)^{-1}$]{GdIr$_2$Si$_2$ ($I4/mmm$) (a) Temperature dependence of the magnetic susceptibility for $B\parallel 100$; (b) Temperature dependence of the magnetic susceptibility for $B=0.01\,\rm T$ along the different crystallographic orientations; (c) Inverse susceptibility for $B=1\,\rm T$. $\mu_{\rm eff}$ and $\Theta_{\rm W}$ have been determined for $B\parallel 001$ and $B\parallel 100$ from the linear fit to the data above $T=200\,\rm K$.}
\label{GdIr2Si2MvT}\label{GdIr2Si2MvT100}\label{GdIr2Si2Chiminus1}
\end{figure*}

\begin{figure*}
\centering
\includegraphics[width=1.0\textwidth]{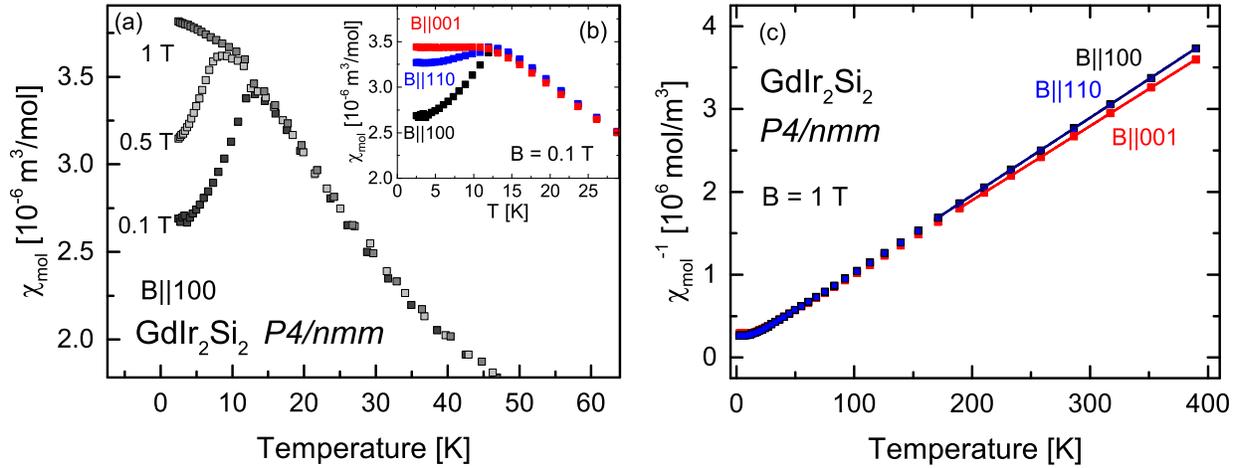}
	\caption[GdIr$_2$Si$_2$ ($P4/nmm$): Susceptibility $\chi(T)$ and inverse susceptibility $\chi(T)^{-1}$]{GdIr$_2$Si$_2$ ($P4/nmm$) (a) Temperature dependence of the magnetic susceptibility for $B\parallel 100$; (b) Temperature dependence of the magnetic susceptibility for $B=0.1\,\rm T$ along the different crystallographic orientations; (c) Inverse susceptibility for $B=1\,\rm T$. $\mu_{\rm eff}$ and $\Theta_{\rm W}$ have been determined for $B\parallel 001$ and $B\parallel 100$ from the linear fit to the data above $T=200\,\rm K$.}
\label{GdIr2Si2_P4_MvT}\label{GdIr2Si2_P4_MvT100}\label{GdIr2Si2_P4_Chiminus1}
\end{figure*}

\subsection{GdIr$_2$Si$_2$: Temperature dependence of the susceptibility}
The Gd$^{3+}$ ions in GdIr$_2$Si$_2$ have a half-filled $4f$ shell and due to the vanishing total orbital momentum $L$, the total angular momentum equals the spin momentum $J=S=7/2$. Above the magnetic ordering temperature, Gd compounds can show a small magnetic anisotropy the cause of which is subject to speculations. As possible origins the dipole interactions, crystal electric field and exchange effects from higher multiplets are discussed \cite{Simon1999, Cochrane1973}.
The temperature dependent susceptibility $\chi(T)$ of GdIr$_2$Si$_2$ with $B \parallel 100$ is presented in Fig.~\ref{GdIr2Si2MvT}(a) ($I4/mmm$) and Fig.~\ref{GdIr2Si2_P4_MvT}(a) ($P4/nmm$). With this field direction, the susceptibility shows strong changes for both compounds below $T_N$. For low fields $B= 0.01\,\rm T$ ($B= 0.1\,\rm T$), $\chi(T)$ Fig.~\ref{GdIr2Si2MvT}(b) ($I4/mmm$) and Fig.~\ref{GdIr2Si2_P4_MvT}(b) ($P4/nmm$) shows an anisotropy for $B\parallel 001$ and $B\perp 001$. 

The antiferromagnetic ordering of the Gd$^{3+}$ moments at $T_N=87\,\rm K $ ($T_N=12\,\rm K $) is indicated for all field directions by a kink in the susceptibility curves.
For the lower symmetric polymorph, the magnetic order is suppressed for $B= 1\,\rm T$ which is not the case for the higher symmetric GdIr$_2$Si$_2$ where a field of $B\approx 2\,\rm T$ is needed to reach the field polarized state.
At high temperatures, $\chi(T)$ exhibits Curie-Weiss behaviour $\chi(T)=C_{\rm mol}/(T-\Theta_{\rm W})$ with the Curie constant $C_{\rm mol}$ and the Weiss temperature $\Theta_{\rm W}$.
For each field direction, we determined the effective magnetic moments from the slope of the inverse susceptibility 
$\chi^{-1}(T)= -\Theta_{\rm W}/C_{\rm mol} + T/C_{\rm mol}$  between $200$ and $390\,\rm K$, according to $\mu_{\rm eff}=\sqrt{(3k_{\rm B} C_{\rm mol})/(\mu_0 N_A)}$, with the Boltzmann constant $k_{\rm B}$, the vacuum permeability $\mu_0$ and the Avogadro number $N_A$. The Weiss temperature was calculated from the intersection of the high-temperature fit, Figs.~\ref{GdIr2Si2MvT}(c) and ~\ref{GdIr2Si2_P4_MvT}(c), with the abscissa for both polymorphous phases.
The effective magnetic moments for GdIr$_2$Si$_2$ I-type phase, $\mu^{\rm 100,110}_{\rm eff} = (8.2\pm 0.2)\,\mu_{\rm B}$ (P-type phase $(8.3\pm 0.2)\,\mu_{\rm B}$) and $\mu^{\rm 001}_{\rm eff} = (8.3\pm 0.2)\,\mu_{\rm B}$ (P-type phase $(8.4\pm 0.2)\,\mu_{\rm B}$),
agree well with the predicted value of $\mu^{\rm calc}_{\rm eff} = 7.94\,\mu_{\rm B}$ for the free Gd$^{3+}$. Slight deviations between the calculated effective magnetic moment and the experimental value of the same order of magnitude were observed already in other LnT$_2$Si$_2$ (Ln$=$rare earth, T$=$ transition metal) compounds, and its origin is not clear so far \cite{Czjzek1989, Kliemt2017}.
The Weiss temperatures for field in and perpendicular to the $a-a$ plane, Tab.~\ref{GdIr2Si2_Theta}, are isotropic for both polymorphs as expected for Gd compounds ($L=0$) due to the absence of crystal electric field effects. While they take values close to $\Theta_{\rm W}=0\,\rm K$ and thus are much smaller than $T_N$ in the I phase, we found a negative Weiss temperature $\Theta_{\rm W}\approx -T{_N}$ indicating the dominance of antiferromagnetic fluctuations for the P phase. 
As demonstrated in Tab.~\ref{GdIr2Si2_Theta}, the T$_N$/$\Theta_W$ ratio changes strongly between the $I$- and the $P$-phase compound. From this, we conclude that there is a strong difference in the $J$ parameters of the Ruderman-Kittel-Kasuya-Yosida (RKKY) interaction. This change of the $J$ parameters is accompanied by a change of the inplane alignment of the magnetic moments. Reliable conclusions whether the change in $J_1$ (Gd-Gd, along $[100]$) or in $J_2$ (along the $[110]$ inplane direction) is larger would require a detailed theoretical analysis.

\begin{figure*}
\centering
\includegraphics[width=1.0\textwidth]{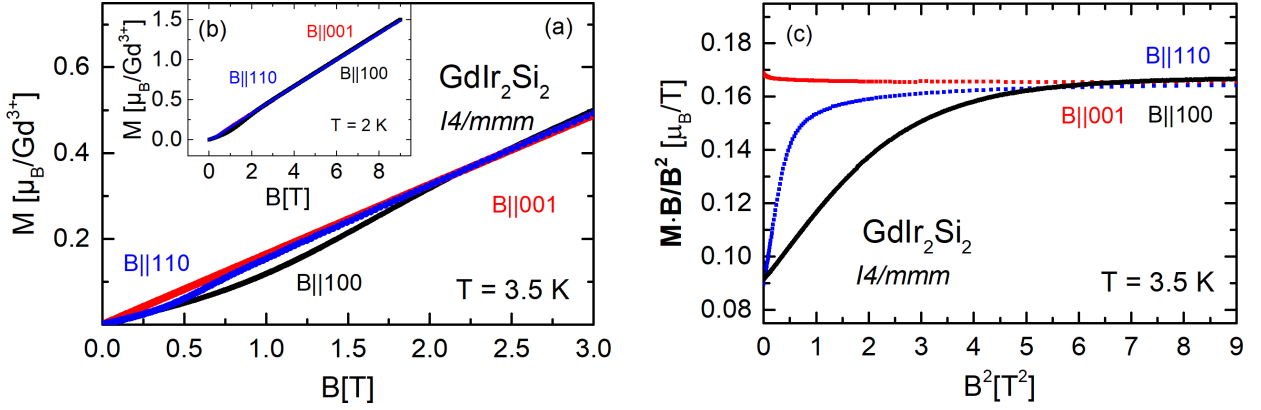}
	\caption[GdIr$_2$Si$_2$ ($I4/mmm$): Magnetization $M(B)$ at $T=3.5\,\rm K$]{GdIr$_2$Si$_2$ ($I4/mmm$) (a) Field dependence of the magnetization for different crystallographic orientations at $T=3.5\,\rm K$ and  (b) at $T=2\,\rm K$ for fields up to $B=9\rm T$; (c) Data in the left panel depicted as $M/B$ versus $B^2$ for $T=3.5\,\rm K$.}
\label{GdIr2Si2MdurchBVsBquad}\label{GdIr2Si2MvH}
\end{figure*}

\begin{figure*}
\centering
\includegraphics[width=1.0\textwidth]{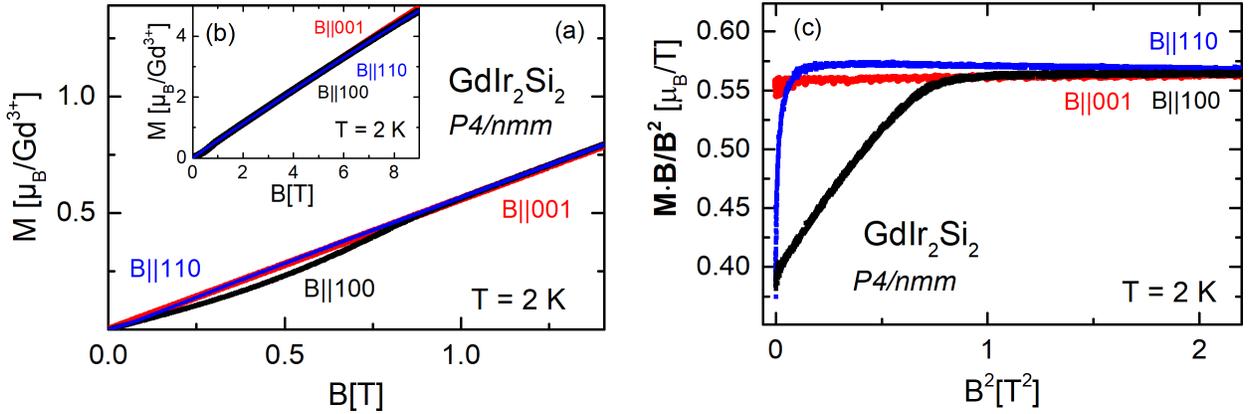}
	\caption[GdIr$_2$Si$_2$ ($P4/nmm$): Magnetization $M(B)$ at $T=2\,\rm K$]{GdIr$_2$Si$_2$ ($P4/nmm$) (a) and (b) Field dependence of the magnetization for different crystallographic orientations at $T=2\,\rm K$; (c) Data of the left panel depicted as $M/B$ versus $B^2$ for $T=2\,\rm K$. }
\label{GdIr2Si2_P4_MdurchBVsBquad}\label{GdIr2Si2_P4_MvH}
\end{figure*}

\subsection{GdIr$_2$Si$_2$: Field dependence of the magnetization}

The field dependence of the magnetization $M(B)$ at $3.5\,\rm K$ of the I-type phase is shown in Fig.~\ref{GdIr2Si2MvH}(a), and the same data are depicted as $\vec{M}\cdot\vec{B}/B^2$ in (c). 
$M(B)$ of the I-type phase shows metamagnetic transitions for $B\parallel 100$ and $B\parallel 110$ for fields below $B=2\,\rm T$.
A different field dependence was observed for the P-type phase, Fig.~\ref{GdIr2Si2_P4_MvH}(a). We found a metamagnetic spin-flop transition only for $B\parallel 100$ below $B=0.9\,\rm T$ at $2\,\rm K$ while the behaviour for $B\parallel 110$ shows the characteristics of a reorientation of magnetic domains similar as it was observed in GdRh$_2$Si$_2$ \cite{Kliemt2017}.
 The susceptibility $\vec{M}\cdot\vec{B}/B^2$ is constant for $B\parallel 001$, $B\leq 9\,\rm T$. For the I-type phase, a constant susceptibility is also reached at $B\approx 2\,\rm T$ ($B\approx 0.2\,\rm T$, P-type phase) for $B\parallel 110$ and at $B\approx 2.5\,\rm T$ ($B\approx 0.9\,\rm T$, P-type phase) for $B\parallel 100$. 
From the fact that the susceptibility $\vec{M}\cdot\vec{B}/B^2$ is constant for $B\parallel 001$, we conclude that the magnetic moments are oriented perpendicular to the crystallographic $[001]$ direction. Recently, this conclusion was confirmed by angle resolved photoemission studies in combination with band structure calculations of the I-type phase \cite{Schulz2021}.
At lower fields, the susceptibility changes strongly with field for both in-plane directions for both polymorphous phases. These observations and the comparison with GdRh$_2$Si$_2$ \cite{Kliemt2015, Kliemt2017} hint to a moment orientation inbetween the $[100]$ and the $[110]$ direction in the $a-a$ plane at $T=3.5\,\rm K$ ($T=2\,\rm K$) in GdIr$_2$Si$_2$, $I4/mmm$ ($P4/nmm$).
In the P-type phase the moments are tilted at most a few degrees away from the $[110]$ direction in the $a-a$ plane and small fields along $B\parallel 110$ already cause a reorientation of magnetic domains.
In both polymorphs, the magnetic moments are aligned in the basal plane of the tetragonal lattice with moments inbetween $[100]$ and $[110]$ in the I-phase compound and moments very close to $[110]$ in the P-phase compound.



\newpage
\begin{table}
\begin{tabular}{|c|c|c|c|c|}
\hline
	Compound&$\Theta_{\rm W}^{100}$&$\Theta_{\rm W}^{110}$&$\Theta_{\rm W}^{001}$&T$_{\rm N}$\\
	&$[\rm K]$             &$[\rm K]$             &$[\rm K]$&     $[\rm K]$       \\
\hline
	GdIr$_2$Si$_2$&            &            &          &    \\
	$I4/mmm$      &$(0\pm 5)$  &$(-3\pm 1)$ &$(5\pm 5)$&86 \cite{Kliemt2018}  \\
	$P4/nmm$      &$(-10\pm 1)$&$(-10\pm 1)$&$(-12\pm 1)$& 12.5 \cite{Kliemt2019} \\
	HoIr$_2$Si$_2$ \cite{Kliemt2018}&&&&\\
	$I4/mmm$      &$(-26\pm 1)$&$(-26\pm 1)$&$(26\pm 1)$&22\\
\hline
	              &$\Theta_{\rm W}^{\rm poly}$ &        &&\\
	              &$[\rm K]$                &           &&\\
\hline
	HoIr$_2$Si$_2$&&&&\\
	$P4/nmm$      &$2\pm 1$        &$-$         &$-$     &5.8  \\
	ErIr$_2$Si$_2$&&&&\\
	$P4/nmm$      &$-5.5\pm 0.3$   &$-$         &$-$& 3.8      \\
\hline 
\end{tabular}
\caption[]{Weiss and N\'eel temperatures of LnIr$_2$Si$_2$ ($I4/mmm$ and $P4/nmm$).}
	\label{GdIr2Si2_Theta}
\end{table}

\newpage


\begin{figure*}
\centering
\includegraphics[width=0.52\textwidth]{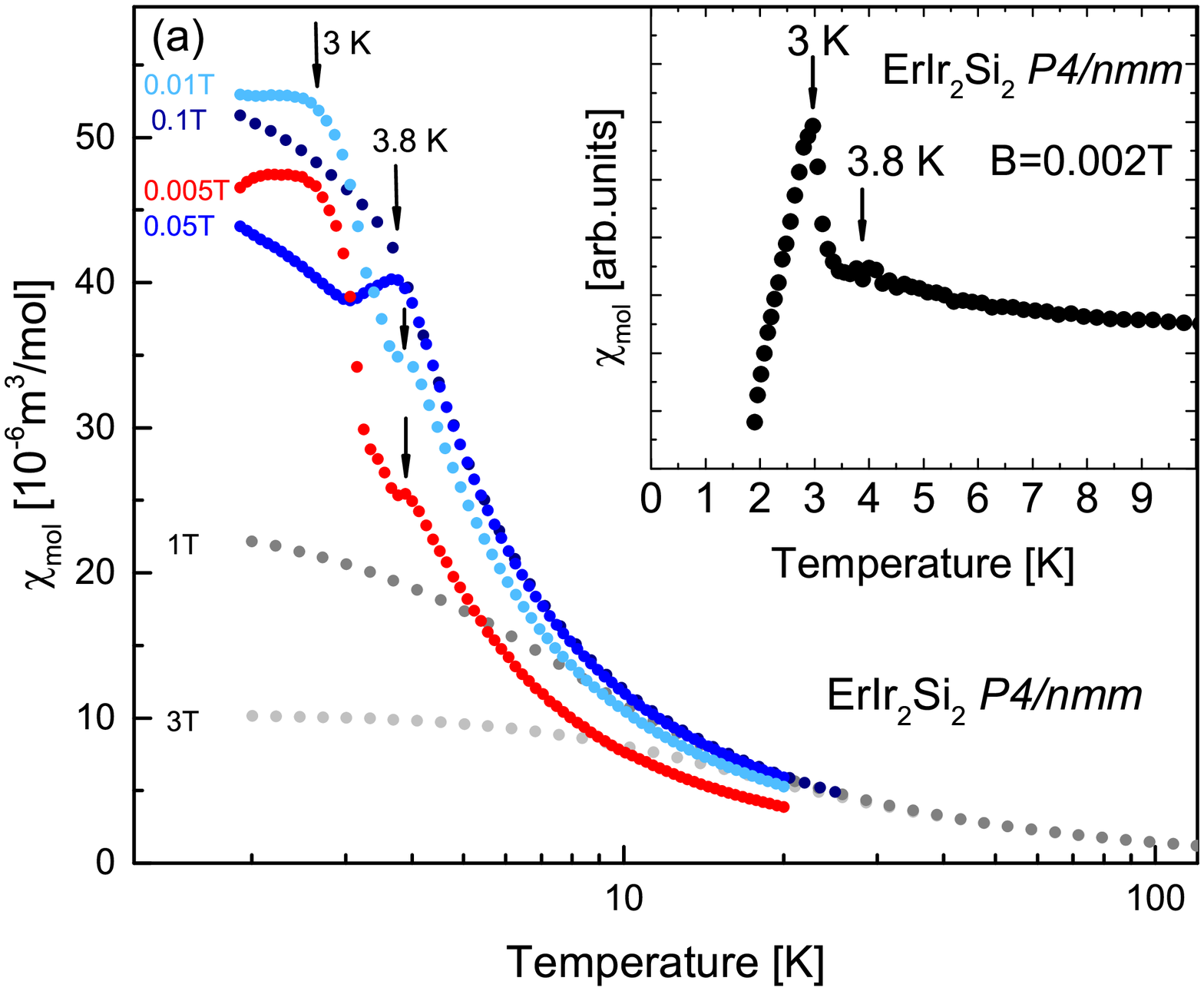}
\includegraphics[width=0.46\textwidth]{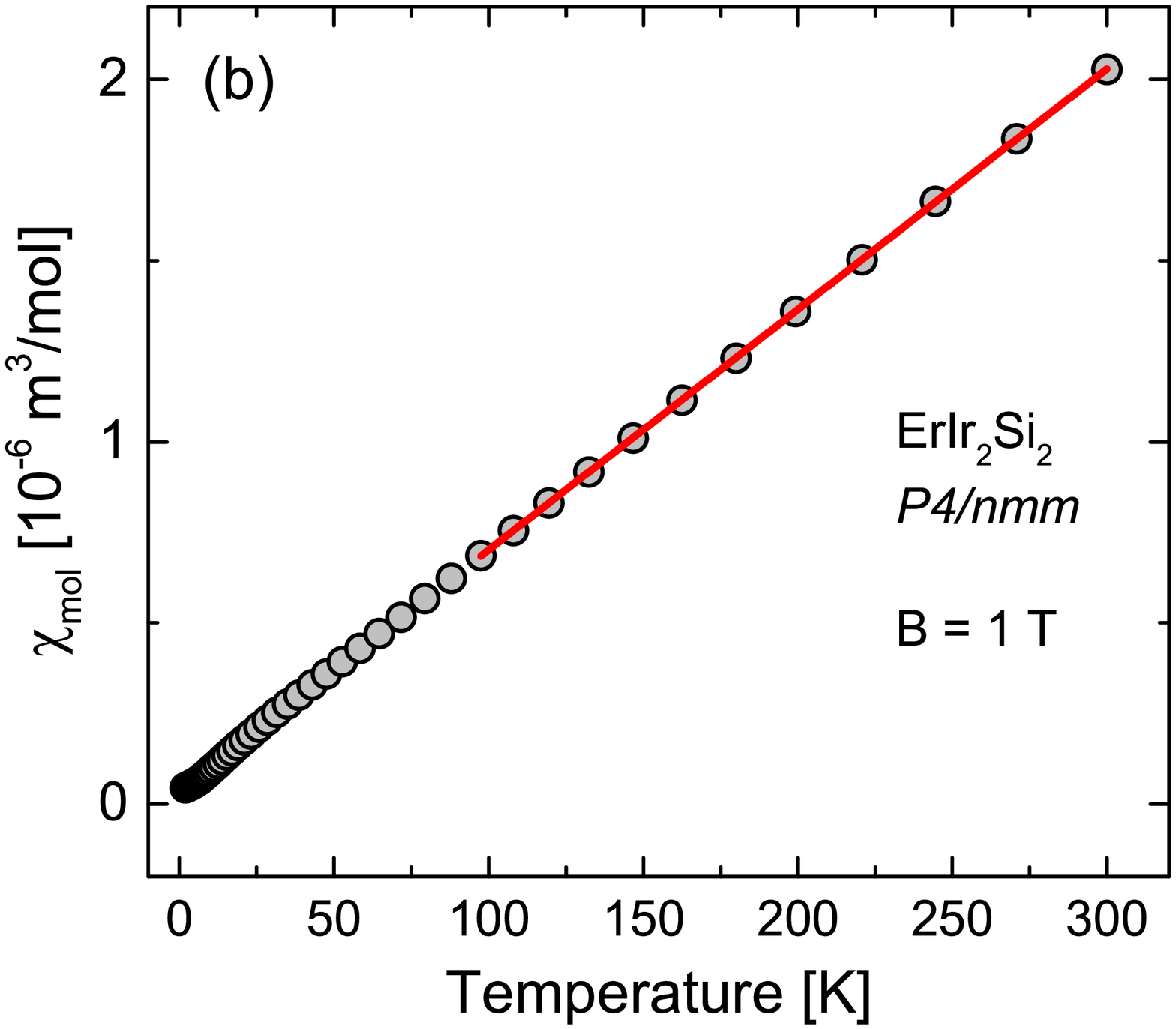}
\caption{ErIr$_2$Si$_2$ ($P4/nmm$): (a) Magnetic susceptibility $\chi(T)$, (b) Inverse susceptibility $\chi(T)^{-1}$. $\mu_{\rm eff}$ and $\Theta_{\rm W}$ have been determined from the linear fit to the data above $T=100\,\rm K$.}
\label{ErIr2Si2MvTbild}
\end{figure*}

\begin{figure*}
\centering
\includegraphics[width=0.49\textwidth]{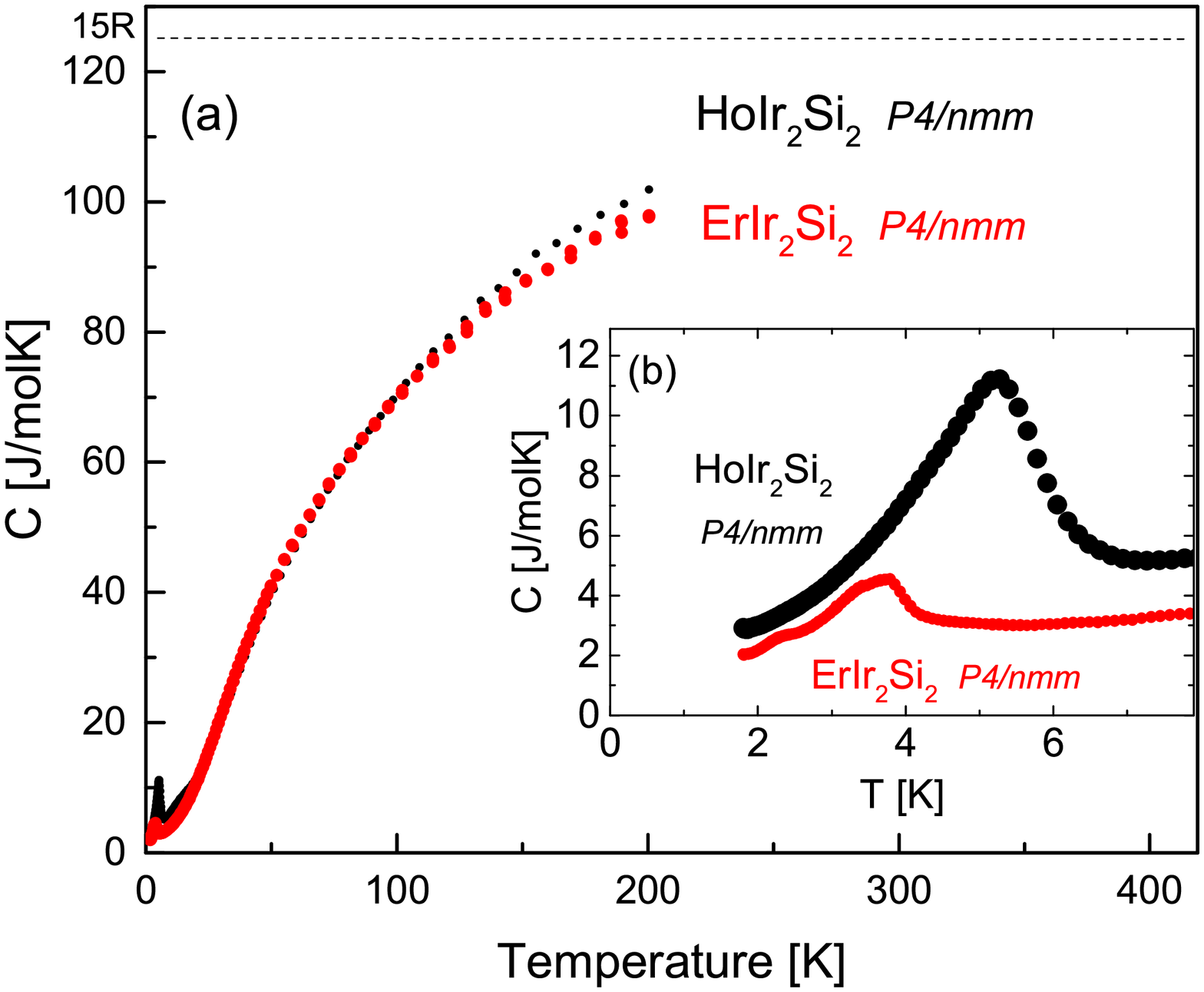}
	\includegraphics[width=0.47\textwidth]{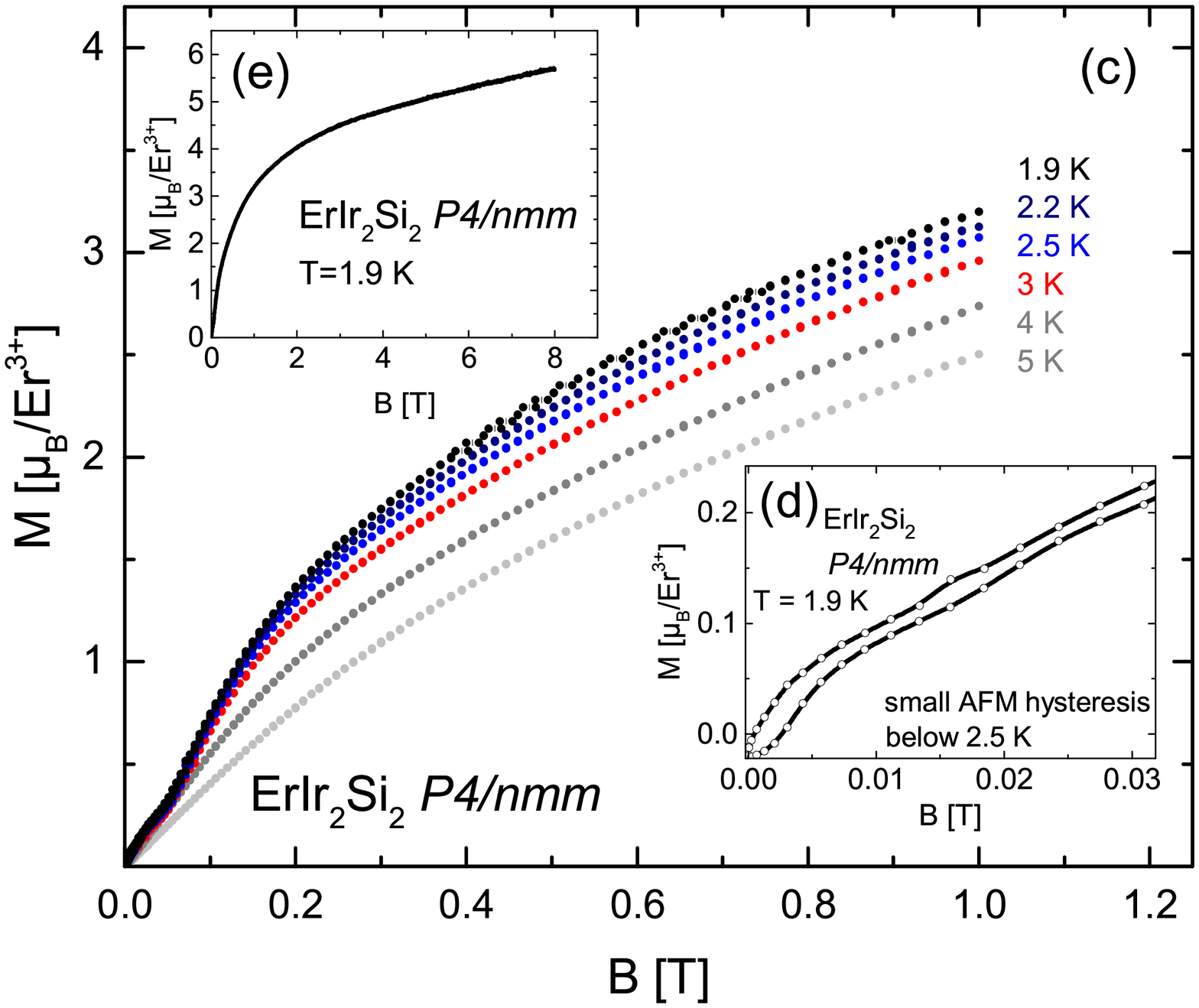}
	\caption[]{HoIr$_2$Si$_2$ ($P4/nmm$, black) and ErIr$_2$Si$_2$ ($P4/nmm$, red) (a) Heat capacity versus temperature, (b) Heat capacity versus temperature between $1.8$ and $8\,\rm K$. ErIr$_2$Si$_2$ ($P4/nmm$): (c) Field dependence of the magnetization, (d) Below $2.5\,\rm K$ a small AFM hysteresis occurs in the M(B) data and (e) Magnetization versus field up to $B = 8\,\rm T$ at $1.9\,\rm K$.}
\label{HoEr_HC}
\end{figure*}

\subsection{ErIr$_2$Si$_2$ thermodynamic and magnetic properties}
The magnetism in ErIr$_2$Si$_2$, $P4/nmm$, is dominated by the local moment of the Er$^{3+}$ ion  
with $S=3/2, L=6$ and $J=15/2$. The magnetization was measured on a polycrystalline sample.
The temperature dependence of the magnetic susceptibility, $\chi(\rm T)$ from $2 - 300\,\rm K$ is shown in Fig.~\ref{ErIr2Si2MvTbild}(a). According to the deGennes scaling for the LnIr$_2$Si$_2$ series \cite{Kliemt2019}, the ordering temperature for ErIr$_2$Si$_2$ $(P4/nmm)$ is expected to be quite low. We observe for $\rm T < 6\,\rm K$ a strong increase of the susceptibility and a peak at $T_{\rm N}\approx 3.8\,\rm K$ where magnetic order sets in. A second peak is visible at $\approx 3\,\rm K$. The magnetic ordering temperature of ErIr$_2$Si$_2$, $I4/mmm$, is $T_{\rm N}=10\,\rm K$ with moments aligned in the basal plane of the tetragonal lattice \cite{Sanchez1993}. 
From the susceptibility data, we conclude, that our sample although it is polycrystalline, probably contains large crystal grains such that main symmetry directions were aligned parallel or perpendicular to the field during the measurement. From the slope of the inverse susceptibility, $\chi^{-1}(\rm T)$ ($B=1\,\rm T$), at high temperatures, Fig.~\ref{ErIr2Si2MvTbild}(b), we determined $\mu_{\rm eff}=(9.8\pm 0.2)\,\mu_{\rm B}$, 
which is close to the expected value of $\mu^{\rm calc}_{\rm eff}=9.58\,\mu_{\rm B}$. The Weiss temperature $\Theta_{\rm W}=-5.50\,\rm K$ has a value close to T$_{\rm N}$, and is negative in this compound.
The field dependence of the magnetization, Fig.~\ref{HoEr_HC}(c), shows a metamagnetic transition below $B\approx 0.1\,\rm T$. In contrast, in polycrystalline 
ErIr$_2$Si$_2$ (I phase) a metamagnetic transition was reported at a much higher field of $B=1.4\,\rm T$ for $T=4.5\,\rm K$ \cite{Sanchez1993}.
At $T=1.9\,\rm K$, $B=8\,\rm T$,  Fig.~\ref{HoEr_HC}(e), $2/3$ of the saturation magnetization $M_{\rm sat}=g_{j}J\mu_{\rm B}=9\,\mu_{\rm B}$ is reached.
The specific heat capacity of ErIr$_2$Si$_2$ ($P4/nmm$) in Fig.~\ref{HoEr_HC}(a),(b) measured down to $1.8\,\rm K$, shows a transition to magnetic order at $T_{\rm N1}=3.8\,\rm K$, a small shoulder at $3\,\rm K$ and a further anomaly at $T_{\rm N2}\approx 2.5\,\rm K$. 
 From the heat capacity data, we deduce a N\'eel temperature of $T_{\rm N}\approx 3.8\,\rm K$ and find further humps at $\approx 2.5\,\rm K$ and $\approx 3\,\rm K$. The inset of Fig.~\ref{ErIr2Si2MvTbild}(a), shows magnetic order at $T{_1}=3\,\rm K$ in $\chi(T)$ with a very low field of $B=0.002\rm T$ aligned along one of the sample edges. With increasing field, the signature at the N\'eel temperature at $T_{\rm N}\approx 3.8\,\rm K$ becomes more pronounced. This hints to a very anisotropic field dependence of the magnetization for the different crystal orientations. We think that, like HoRh$_2$Si$_2$, this material might exhibit a "component-separated magnetic order" \cite{Shigeoka2012} meaning that different ordering temperatures can be found for field parallel and perpendicular $c$. 
We subtracted the specific heat of the non-magnetic reference LaIr$_2$Si$_2$ \cite{Valiska2012} from the heat capacity data of ErIr$_2$Si$_2$ and HoIr$_2$Si$_2$ to determine the involved entropy at the magnetic phase transition. According to Kramers theorem, the CEF ground state of ErIr$_2$Si$_2$ which possesses an odd number of $4f$ electrons is at least a doublet and the entropy involved in the phase transition in this case is S$^{4f}= \rm Rln2$. For ErIr$_2$Si$_2$ we found that S$^{4f}= \rm Rln2$ is reached near $10\,\rm K$ which hints to a doublet CEF ground state in this material. For HoIr$_2$Si$_2$ we can exclude a doublet as the ground state since S$^{4f}$  reached about $1.7\,\rm Rln2$ at $10\,\rm K$.


\subsection{HoIr$_2$Si$_2$ thermodynamic and magnetic properties}

\begin{figure*}
\centering
\includegraphics[width=1.0\textwidth]{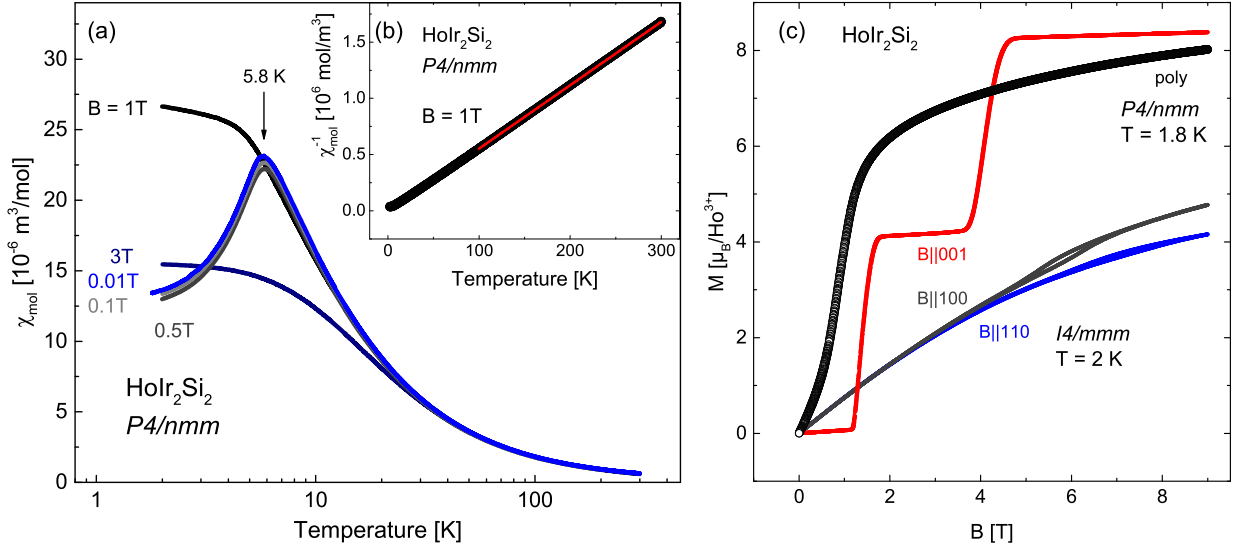}
	\caption[]{HoIr$_2$Si$_2$ ($P4/nmm$) (a) Magnetic susceptibility $\chi(T)$, (b) Inverse susceptibility $\chi(T)^{-1}$. The effective magnetic moments $\mu_{\rm eff}$ and Weiss temperatures $\Theta_{\rm W}$ have been determined from the linear fit to the data above $T=100\,\rm K$. (c) Field dependence of the magnetization at $1.8\,\rm K$ of polycrystalline material (P phase, open symbols) and a single crystal (I phase, grey, blue and red lines) at $2\,\rm K$ \cite{Kliemt2018}.}
\label{HoIr2Si2MvT}
\end{figure*}

The specific heat of HoIr$_2$Si$_2$ ($P4/nmm$) is shown in Fig.~\ref{HoEr_HC}(a). At $200\,\rm K$, about $80\%$ of the Dulong-Petit value of $15\,\rm R=125\,\rm J/molK$ is obtained. The compound orders magnetically at $T_{\rm N}= 5.8\,\rm K$, Fig.~\ref{HoEr_HC}(b). 
In HoIr$_2$Si$_2$, $P4/nmm$, we found local moment magnetism of the Ho$^{3+}$ ion  
($S=2,\, L=6$ and $J=8$).
The magnetic susceptibility, $\chi(\rm T)$ of HoIr$_2$Si$_2$ ($P4/nmm$) from $2 - 300\,\rm K$ shows Curie-Weiss behaviour and is presented in Fig.~\ref{HoIr2Si2MvT}(a). While the P phase orders antiferromagnetically at $\rm T_N = 5.8\,\rm K$, $T_{\rm N}=22\,\rm K$ was found for the I phase \cite{Kliemt2018}. In the P phase, the magnetic order is suppressed for $B\geq 1\,\rm T$ but at a much higher field of $B\approx 4\,\rm T$ in the I phase \cite{Kliemt2018}. 
The inverse susceptibility $\chi^{-1}(\rm T)$ ($B=1\,\rm T$) is linear between $300 - 100\,\rm K$ as it is shown in the inset, Fig.~\ref{HoIr2Si2MvT}(b). From the slope of a linear fit we determined $\mu_{\rm eff}=(10.6\pm 0.2)\mu_{\rm B}$,
which is close to the expected value of $\mu^{\rm calc}_{\rm eff}=10.61\mu_{\rm B}$. The positive Weiss temperature $\Theta_{\rm W}=(1.9\pm 0.2)\rm K$ is close to T$_N$ in this material.
The field dependence of the magnetization, Fig.~\ref{HoIr2Si2MvT}(c), measured at $\rm T = 2\,\rm K$ shows a metamagnetic transition at $B \approx 1\,\rm T$.
The saturation magnetization is M$_{\rm sat}=g_{j}J\mu_{\rm B}=10\mu_{\rm B}$ and about 80\% of this value is reached at about $B=9\rm\, T$.
The behaviour in field of HoIr$_2$Si$_2$ ($P4/nmm$) with only one field induced metamagnetic transition at $B \approx 1\,\rm T$ is in contrast to that of HoIr$_2$Si$_2$ ($I4/mmm$). The later shows two transitions in field parallel to the $(001)$ direction for $B_{1} \approx 1.5\,\rm T$ and $B_{2} \approx 4\,\rm T$ \cite{Kliemt2018}.

\section{Summary}
In our study, we present a magnetic characterization of two different tetragonal polymorphs of GdIr$_2$Si$_2$.
The single crystals have been grown in the I-type phase (space group $I4/mmm$) and in the P-type phase (space group $P4/nmm$) at high temperatures between $T_{\rm max}=1550^{\circ}$C and $1090^{\circ}$C by a modified Bridgman method from indium flux. After an optimization of the temperature-time profile and the initial stoichiometry of the growth experiment, we obtained millimetre-sized single crystals with a platelet habitus with the $c$-axis perpendicular to the platelets in the $I4/mmm$ and the $P4/nmm$ tetragonal structure. 
The temperature dependent susceptibility indicates an antiferromagnetic ordering temperature of $T_{\rm N}=87\,\rm K$ for crystals in the I-type phase. The P-type phase with a reduced crystal symmetry compared to the I phase shows a much lower N\'eel temperature of $T_{\rm N}=12\,\rm K$. 
The effective magnetic moments for the different field directions are close to the effective magnetic moment, $\mu_{\rm eff}^{\rm theo}=7.94\,\mu_B$, of a free Gd$^{3+}$ ion. We found different behaviour concerning the high-temperature fluctuations in the two polymorphs. For the I-type phase, the Weiss temperatures take values close to $\Theta_{\rm W}=(0\pm 5)\,\rm K$ and thus are much smaller than $T_N$, while we determined $\Theta_{\rm W}\approx -T{_N}$ for the P-type phase. The negative Weiss temperature indicates the dominance of antiferromagnetic fluctuations in the P-type phase. 
The field dependence of the magnetization hints to a moment orientation in the basal plane of the tetragonal lattice for both structural polymorphs. The reduction of the symmetry of the crystal structure from $I4/mmm$ to $P4/nmm$ leads to changes in the band structure, namely a loss of degeneracy of certain bands and therefore increases the number of bands which apparently influences the magnetic properties of the polymorphs. In the I-type phase, the moments are aligned inbetween the $[100]$ and $[110]$ whereas the P-type phase shows moment orientation very close to the $[110]$ direction. 
We prepared polycrystalline samples of HoIr$_2$Si$_2$ ($P4/nmm$) and found that this compound orders magnetically below $T_{\rm N}=5.8\,\rm K$. The field dependent magnetization shows a metamagnetic transition below $B \approx 1\,\rm T$. Furthermore, we confirmed the existence of ErIr$_2$Si$_2$ in the space group $P4/nmm$. In the magnetic susceptibility we found a strong increase below $6\,\rm K$ as sign for the onset of magnetic order. From the heat capacity we determined the ordering temperature $T_{\rm N}=3.8\,\rm K$. A change of the magnetic structure might occur at $\approx 2.5\,\rm K$ where a small hump in the heat capacity is visible. The polymorphs of GdIr$_2$Si$_2$, HoIr$_2$Si$_2$ and ErIr$_2$Si$_2$ show the same trend as all other up to now known LnIr$_2$Si$_2$ compounds: The lower symmetric $P4/nmm$ polymorph orders magnetically at a lower temperature than the higher symmetric $I4/mmm$ compound. For the LnIr$_2$Si$_2$ polymorphs that were studied, we found that a much higher field is needed in the I-phase compounds compared to the P-phase compounds to suppress the magnetic order.

\cleardoublepage

\begin{table}
\begin{tabular}{|c|c|c|c|c|c|c|}
\hline
Compound&Structure &$T_{\rm N}$&$a$&$c$&$c/a$                             &Ref.\\
        &type      &               [K]   &        $[$\AA$]$   & $[$\AA$]$     &          & \\   
		 \hline
 CeIr2Si2 &$I4/mmm$             &PM$^*$ to 2\,K  &4.086&        10.164&        2.49&\cite{Mihalik2009, Zhong1985}\\        
               &$P4/nmm$            &PM$^*$ to 2\,K &4.147&        9.881&        2.38&\cite{Mihalik2009, Zhong1985}\\        
PrIr2Si2&         $I4/mmm$           &45.5,  23.7          &        4.084(2)&        10.145(2)&2.484(3)&\cite{Mihalik2010}\\
             &$P4/nmm$           &PM$^*$ bis 2\,K          &4.152      &9.847        &2.37&\cite{Mihalik2010}\\
NdIr2Si2 & $I4/mmm$          &33, 18                      &4.077(1) &10.099(3)  &2.477(4)&\cite{Welter2003}\\
                &$P4/nmm$          &            $-$              &4.144       &9.851          &2.38       &\cite{Zhong1985}\\
SmIr2Si2&         $I4/mmm$        &38.9                    &4.063       &10.016        &2.47       &\cite{Valiska2013}\\
                &$P4/nmm$        &22                       &4.124       &9.806           &2.38&\cite{Valiska2013}\\
	EuIr2Si2 & $I4/mmm$        &PM$^*$, iv$^{**}$  &4.083 &10.123&2.48&\cite{Chevalier1986}\\      
GdIr2Si2 &$I4/mmm$        &86                     &4.06&        9.93& 2.45&\cite{Kliemt2019}\\
                &$P4/nmm$        &12.5&4.111&9.733  &2.37&\cite{Kliemt2019}\\
	TbIr2Si2 & $I4/mmm$        &72&4.048(5)   &9.911(5)&2.45&\cite{Hirjak1984}\\
                &$P4/nmm$        &11.5, 5        &4.101        &9.643&2.35&\cite{Shigeoka2015}\\
DyIr2Si2 & $I4/mmm$        &40          &4.029       &9.856&2.45&\cite{Sanchez1993, Melamud1984}\\
                &$I4/mmm$        &30,17,9        &4.05        &9.75&2.41&\cite{Uchima2017}\\
                &$P4/nmm$        &9.4,3.0        &4.10        &9.67&2.36&\cite{Uchima2017}\\
HoIr2Si2&         $I4/mmm$        &22           &4.048       &9.884&2.44&\cite{Kliemt2018}\\
                &$P4/nmm$        &  $-$             &4.085       &9.572&2.34&\cite{Zhong1985}\\
                &$P4/nmm$        &5.8        &4.0796      &9.6677&2.37&this work, with\\
			&&&&&&14\% I phase\\

ErIr2Si2 &$I4/mmm$        &10         &4.028(2)   &9.855(9)&2.45&\cite{Slaski1982, Sanchez1993}\\
                &$P4/nmm$  &    3.8     &4.0813        &9.6261&2.36&this work, with \\
			&&&&&&27\% I phase\\

		 &$P4/nmm$   &    $-$     &4.0666(13)&9.583(4)&2.36&this work,\\
		 &&&&&& SC analysis\\
YbIr2Si2 & $I4/mmm$&      $-$        &4.0345(15)&9.828(2)&2.44&\cite{Hossain2006}\\
                 &$P4/nmm$&    0.6          &4.0370(20)&9.898(5)&2.45&\cite{Hossain2006, Krellner2012a}\\
\hline
\end{tabular}
	\caption[]{Lattice parameters and magnetic ordering temperatures of the LnIr$_2$Si$_2$ ($I4/mmm$ and $P4/nmm$) compounds. $^*$ paramagnetic, $^{**}$ intermediate valent}
	\label{LnIr2Si2_magnet_lattice}
\end{table}

\cleardoublepage
\section*{Acknowledgments}
We acknowledge discussions with S.Schulz, M.Otrokov, D.V.Vyalikh and C.Geibel as well as funding by the Deutsche Forschungsgemeinschaft (DFG, German Research Foundation) through grant No. KR3831/5-1 and via the TRR 288 (422213477, project A03).


\section*{References}






\bibliography{diss_LnIr2Si2}





\end{document}